\documentclass[twocolumn,prb,aps,showpacs]{revtex4}
\usepackage{graphicx}
\usepackage{bm}
\usepackage{amsmath}
\begin{document}
\title{Temperature and size-dependent suppression of Auger recombination in quantum-confined lead salt nanowires}
\author{Valery I. Rupasov}
\email{valery_rupasov@hotmail.com}
\affiliation{ANTEOS, Inc., Shrewsbury, Massachusetts 01545, USA}

\begin{abstract}
Auger recombination (AR) of the ground biexciton state in quantum-confined lead salt nanowires (NWs) with a
strong coupling between the conduction and the valence bands is shown to be strongly suppressed, and only
excited biexciton states contribute to Auger decay. The AR rate is predicted to be greatly reduced when
temperature or the NW radius are decreased, and the effect is explained by decrease in both the population
of excited biexciton states and overlap of phonon-broadened single- and biexciton states. Suppression
of AR of multiexciton states exhibiting strong radiative decay makes obviously lead salt NWs a subject
of special interest for numerous lasing applications.
\end{abstract}

\pacs{73.21.Hb, 73.22.Dj, 78.67.Lt}

\maketitle

Auger recombination (AR) is a nonradiative process, in which the electron-hole recombination energy is
transferred to a third charge carrier \cite{L} exciting it to a higher-energy state. In bulk semiconductors,
AR is inefficient because of constrains imposed by energy and translational-momentum conservation. But
due to relaxation of translation-momentum conservation, the efficiency of Auger-like processes is greatly
enhanced in quantum-confined nanocrystals (NCs), \cite{CE,K1,K2} where AR plays a central role in
carrier relaxation.

Effectively one-dimensional nanowires (NWs) occupy an intermediate position between zero-dimensional NCs
and bulk materials. In sufficiently long NWs of the length $L$ essentially exceeding the NW radius $R$,
a longitudinal motion of charge carriers along the NW axis is free, while a transverse motion in the plane
perpendicular to the NW axis is spatially quantized. The electronic structure of quantum-confined NWs is
composed of subbands of the longitudinal motion with a wave vector $k$, which are also characterized by
a spatially quantized wave vector of the transverse motion $p$. Since electronic transitions between
subbands with different $p$ are not forbidden by longitudinal wave vector conservation, quantum confinement
in NWs is expected to result in enhancement of Auger effect compared with bulk parent materials. In particular,
Auger decay of the ground biexciton state, \cite{N} which is forbidden in the bulk by energy and
translation-momentum conservation, is expected to be allowed in quantum-confined NWs.

A strong coupling between the conduction and the valence bands in lead salt materials results in significant
corrections to the electronic structures of NCs \cite{KW} and NWs \cite{R} computed in the framework of
``particle-in-a-box'' model  \cite{ER} not accounting for the interband coupling. However, in the case of
NCs, these corrections are mainly quantitative and do not lead to new qualitative physical results. In
particular, no essential difference is observed in size dependence of AR in narrow-gap NCs and NCs made of
wide-gap materials, \cite{K2} in which contribution of the interband coupling is relatively small.

But a situation is essentially changed for NWs with a strong interband coupling. In contrast to the case of
NCs, the interband coupling in NWs significantly lifts degeneration of low-energy single-particle states of
the transverse electron motion resulting in specific selection rules for Auger-like processes. That in its
turn leads to qualitative changes in temperature and size behavior of AR compared with those in NCs and bulk
materials.

In this paper, employing results of our recent studies \cite{R} of the electronic structure of quantum-confined
lead salt NWs, we show that, in sharp contrast to the case of NCs, AR of the ground biexciton state is strongly
suppressed, and only excited biexciton states, which are partially populated at finite temperature, contribute
to Auger decay of biexcitons. Thus, the total rate of biexciton AR is given by the sum of the rates of excited
biexciton states only with conventional weight factors, which determine their population at finite temperature $T$:
\begin{equation}
W(T,R) = \sum_n W_n(R)e^{-\Delta E_n(R)/T}.
\end{equation}
Here, size-dependent $\Delta E_n$ is the energy difference between $n$-th excited biexciton state with
the AR rate $W_n$ and the ground biexciton state. Analytical and numerical computations show that, again in
contrast to the case of NCs, size dependence of matrix elements of effective Coulomb coupling for Auger-like
processes in NWs is very slow, and size dependence of the rates $W_n$ is mainly governed by size-dependent
overlap of phonon-broadened single- and biexciton states.

Thus, despite strong spatial quantization of the transverse electron motion resulting in significant relaxation
of translation-momentum conservation, temperature and size behavior of AR in NWs qualitatively differs from that
in NCs. However, this effect is observed only in NWs with a strong interband coupling, and it is not described
in the framework of particle-in-a-box model.

If carrier motions in the conduction and the valence bands are treated as independent of each other,
envelope electronic wave functions in the cylindrical coordinate system $(r,\phi,z)\equiv(\bm{r},z)$
with the $Z$ axis directed along the NW axis are easily found as
$\chi_{m_l,s_z}=J_{m_l}(pr)e^{im_l\phi}e^{ikz}\sigma_a$, where $J$ is the Bessel function, and $\sigma_a$
($a = \uparrow,\downarrow$) are conventional spinors
$
\sigma_\uparrow = \left(\begin{array}{c}1\\0\end{array}\right),\;\;\;
\sigma_\downarrow = \left(\begin{array}{c}0\\1\end{array}\right)
$. Spatially quantized wave vectors of the transverse motion $p_n$ are found from the boundary condition on the
NW interface $J_{m_l}(r=R) = 0$. The quantum states are characterized by the orbital angular momentum projection
of the transverse motion on the $Z$ axis $m_l=0,\pm 1, \ldots$, the electron spin projection $s_z=\pm\frac{1}{2}$,
and by continuous wave vector of the longitudinal motion $k$. Due to mirror symmetry of the conduction and the
valence bands, the subband energies $E$ are found to be $\pm\epsilon_{p,k}$ for the conduction ($+$) and the valence
($-$) bands, where $\epsilon_{p,k} = \frac{1}{2}E_{\text{g}}+\frac{\hbar^2}{2m}(p^2+k^2)$, $E_{\text{g}}$ is the
energy gap, and $m$ is the effective electron mass in both the conduction and the valence bands. The subband states
are degenerate with respect to the sign of the orbital angular momentum projection $m_l$, two possible directions of
the electron spin, and two possible directions of the wave vector $k$. Auger decay of the ground biexciton state is
allowed, and the AR rate does not vanish at $T=0$.

In the four-band envelope-function formalism, taking into account a strong coupling between the conduction
and the valence bands, total electronic wave functions in lead-salt NCs and NWs are written as a product,
$\psi = \sum_{i=1}^{i=4}{\cal F}_iu_i$, of the four band-edge Bloch functions $u_i$ of the conduction ($i=1,2$)
and the valence (i=3,4) bands, and four-component envelope functions ${\cal F}_i$. The transverse electron
motion in NWs is described by bispinors
\begin{subequations}
\begin{equation}
\Psi_{\pm,m_j}=\frac{1}{\sqrt{4\pi\varepsilon_p}}\left(
\begin{array}{c}
  \sqrt{\varepsilon_p\pm\epsilon_p}f_{m_l}(r)e^{im_l\phi}\sigma_\uparrow \\
  \pm\sqrt{\varepsilon_p\mp\epsilon_p}f_{m'_l}(r)e^{im'_l\phi}\sigma_\downarrow\\
\end{array}\right)
\end{equation}
and
\begin{equation}
\Phi_{\pm,m_j}=\frac{1}{\sqrt{4\pi\varepsilon_p}}\left(
\begin{array}{c}
  \sqrt{\varepsilon_p\pm\epsilon_p}f_{m'_l}(r)e^{im'_l\phi}\sigma_\downarrow\\
 \pm\sqrt{\varepsilon_p\mp\epsilon_p}f_{m_l}(r)e^{im_l\phi}\sigma_\uparrow\\
\end{array}\right),
\end{equation}
\end{subequations}
where $m_l=m_j-\frac{1}{2}$, $m'_l=m_j+\frac{1}{2}$, $\epsilon_p=\epsilon_{p,k=0}$,
$\varepsilon_p=\sqrt{\epsilon_p^2+\eta^2p^2}$ is the subband-edge energy, $\eta$ is a parameter
of the interband coupling, and $f_{m_l}(r)$ are radial wave functions \cite{R} normalized to unity.
The total electronic structure is described by bispinors
\begin{subequations}
\begin{equation}
F_{\pm,m_j} =\frac{1}{\sqrt{2E}}\left(\sqrt{E\pm\varepsilon}\Psi_{+,m_j}
\mp\sqrt{E\mp\varepsilon}\Phi_{-,m_j}\right)\frac{e^{ikz}}{\sqrt{L}}
\end{equation}
and
\begin{equation}
G_{\pm,m_j}=\frac{1}{\sqrt{2E}}\left(\sqrt{E\mp\varepsilon}\Psi_{-,m_j}\pm
\sqrt{E\pm\varepsilon}\Phi_{+,m_j}\right)\frac{e^{ikz}}{\sqrt{L}}
\end{equation}
\end{subequations}
in the conduction ($+$) band with the eigenenergy $E=+\sqrt{\epsilon_{p,k}^2+\eta^2(p^2+k^2)}$,
and in the valence ($-$) band with the eigenenergy $-E$, where $\varepsilon=\sqrt{\epsilon_{p,k}^2+\eta^2p^2}$.
Electronic subbands are characterized by the total angular momentum projection
$m_j=m_l+s_z=\pm\frac{1}{2},\pm\frac{3}{2},\ldots$ on the $Z$ axis and the wave vector of longitudinal motion $k$.
\begin{figure}[t]
\includegraphics[width=0.7\linewidth]{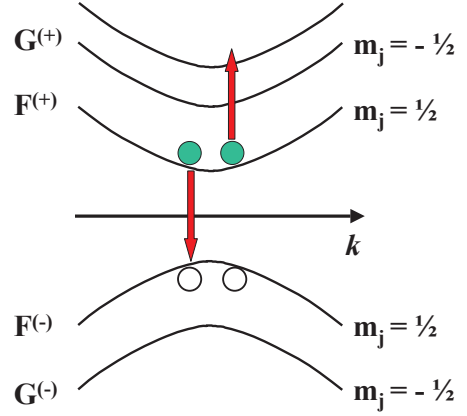}
\caption{Subbands of the longitudinal electron motion in NWs with the strong interband coupling.}
\end{figure}

Two distinct sets of spatially quantized wave vectors of the transverse motion $p_n$ and $q_n$
($n=0,1, \ldots$) are found from boundary condition equations
\begin{subequations}
\begin{equation}
\frac{\sqrt{\varepsilon_p+\epsilon_p}}{\sqrt{\epsilon_\lambda-\varepsilon_p}}
\frac{J_{m_j-\frac{1}{2}}(pR)}{I_{m_j-\frac{1}{2}}(\lambda R)}=
\frac{\sqrt{\varepsilon_p-\epsilon_p}}{\sqrt{\epsilon_\lambda+\varepsilon_p}}
\frac{J_{m_j+\frac{1}{2}}(pR)}{I_{m_j+\frac{1}{2}}(\lambda R)}
\end{equation}
for the bispinors $\Psi_+$ and $\Phi_-$, and
\begin{equation}
\frac{\sqrt{\varepsilon_p+\epsilon_q}}{\sqrt{\epsilon_\lambda-\varepsilon_p}}
\frac{J_{m_j+\frac{1}{2}}(qR)}{I_{m_j+\frac{1}{2}}(\lambda R)}=
\frac{\sqrt{\varepsilon_p-\epsilon_q}}{\sqrt{\epsilon_\lambda+\varepsilon_p}}
\frac{J_{m_j-\frac{1}{2}}(qR)}{I_{m_j-\frac{1}{2}}(\lambda R)}
\end{equation}
\end{subequations}
for the bispinors $\Psi_-$ and $\Phi_+$, where $\lambda=\sqrt{p^2+\lambda_0^2}$,
$\lambda_0=\frac{2m}{\hbar^2}\sqrt{\frac{\hbar^2}{2m}E_{\text{g}}+\eta^2}$, and
$\epsilon_\lambda=\epsilon_{p,k=0}+\frac{2m}{\hbar^2}\eta^2$. Since the longitudinal motion mixes $\Psi_+$-
and $\Phi_-$-bispinors in bispinors $F$, and $\Psi_-$- and $\Phi_+$-bispinors in bispinors $G$, Eqs. (4a)
and (4b) determine spatially quantized wave vectors $p_n$ and $q_n$ for $F$- and $G$-subbands, respectively.
Thus, the interband coupling completely lifts degeneration of transverse motion states, and subband states
in a given $L$-valley are degenerate with respect to two possible directions of the wave vector $k$ only.

In terms of envelope functions, a matrix element of long-range Auger process in the conduction band for
the ground biexciton state, shown in Fig. 1, is derived as
\begin{eqnarray}
\Gamma_\text{g}&\!\!\!\!\!\!\!=\!\!\!\!\!\!\!&\int\!\!d\bm{r}_1 d\bm{r}_2 [F^\dag_{-,\frac{1}{2}}(\bm{r}_1){\cal F}^\dag_{+,m_j}(\bm{r}_2)
\!-\!{\cal F}^\dag_{+,m_j}(\bm{r}_1)F^\dag_{-,\frac{1}{2}}(\bm{r}_2)]\nonumber\\
&&\times {\cal U}(\bm{r}_1,\bm{r}_2)F_{+,\frac{1}{2}}(\bm{r}_2)F_{+,\frac{1}{2}}(\bm{r}_1),
\end{eqnarray}
where the wave functions of the longitudinal motion $\frac{1}{\sqrt{L}}e^{ikz}$ are included into
an effective Coulomb coupling
\begin{eqnarray*}
{\cal U}(\bm{r}_1,\bm{r}_2)&=&\frac{1}{L^2}\int_0^L dz_1dz_2U(\bm{r}_1,z_1;\bm{r}_2,z_2)\nonumber\\
&&\times\exp\{i[(k_1-k'_1)z_1+(k_2-k'_2)z_2]\},
\end{eqnarray*}
where $U$ is the energy of both direct Coulomb coupling and interparticle coupling via medium polarization.
Owing to conservation of the total angular momentum projection, an electron in the conduction band can be
excited only to subbands with $m_j=\frac{1}{2}$.

Nearby subband edges, where the energy differences $E-\varepsilon$ in Eqs. (3) are small, the expressions
for $F$- and $G$-bispinors are simplified:
\begin{eqnarray*}
F_+&\simeq &\Psi_+-\rho\Phi_-;\;\;\;\; F_-\simeq\rho\Psi_+ +\Phi_-\\
G_+&\simeq &\rho\Psi_- +\Phi_+;\;\;\;\; G_-\simeq \Psi_- -\rho\Phi_+,
\end{eqnarray*}
where $\rho=\frac{\eta k}{2\varepsilon_0}$ is the ratio of the longitudinal interband coupling energy to
the energy of the transverse motion. In PbSe material, the parameter of interband coupling $\eta\simeq 0.31$
eV$\cdot$nm, and, at small $k\sim 2\pi/L$ and $L\sim 10$ $\mu$m, the parameter $\rho$ is estimated to
be of the order of $(2-5)\times 10^{-4}$ for PbSe NWs of the radius of $2-8$ nm.

Due to orthogonality of the spinors $\sigma_a$, $\sigma^\dag_\uparrow\cdot\sigma_\downarrow = 0$, $\Psi$- and
$\Phi$-bispinors are also orthogonal to each other, $\Psi^\dag_{m_j}\cdot\Phi_{m'_j}=0$. Therefore, a contribution
of recombination process for an electron from $F_+$-subband and a hole from $F_-$-subband to the matrix element
$\Gamma_\text{g}$ is proportional to $\rho$. Correspondingly, the contribution to the AR rate
$W_\text{g}\sim\Gamma_\text{g}^2$ is of the order of $\rho^2\sim 10^{-7}$, i.e., AR of the ground biexciton state
is strongly suppressed.

In general, we derive the following selection rules for Auger-like processes valid for states with sufficiently
small wave vectors of the longitudinal motion:
\begin{itemize}
\item electron-hole recombination is allowed (i.e., it does not bring the small factor $\rho$ into a matrix
element of Auger-like process) only if an electron and a hole belong to different subbands: $F_+$ and $G_-$,
or $G_+$ and $F_-$;
\item on the contrary, electronic intraband transitions are allowed only inside $F$- and $G$-manifolds, since
a transition, in which an electron is transferred from $F$($G$)- to $G$($F$)-subband inside the conduction
(valence) band, brings the small factor $\rho$ into a matrix element of the process.
\end{itemize}
\begin{figure}[t]
\includegraphics[width=0.7\linewidth]{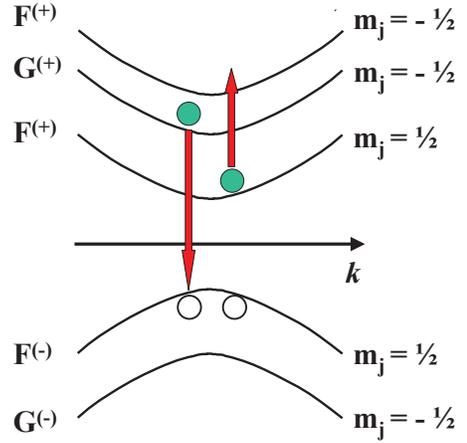}
\caption{Auger recombination of the first excited biexciton state, in which one of electrons is in the subband
$G_{+,-\frac{1}{2}}$.}
\end{figure}

The first excited biexciton states are constructed from the ground state, if one of electrons (holes) is
transferred to the subband $G_{+,-\frac{1}{2}}$ ($G_{-,-\frac{1}{2}}$), and the expression for matrix element
of Auger decay of the excited biexciton state illustrated in Fig. 2 is derived as
\begin{eqnarray}
\Gamma_n&=&\int d\bm{r}_1 d\bm{r}_2 F^\dag_{-,\frac{1}{2}}(\bm{r}_1)F^\dag_{+,-\frac{1}{2}}(\bm{r}_2,p_n)
{\cal U}(\bm{r}_1,\bm{r}_2)\nonumber\\
&&\times F_{+,\frac{1}{2}}(\bm{r}_2)G_{+,-\frac{1}{2}}(\bm{r}_1),
\end{eqnarray}
where ${\cal U}$ does not depend on $R$ at small $\Delta k = |k_1-k'_1|=|k_2-k'_2|\ll R^{-1}$, and $n=1,2,\ldots$
numerates the spatially quantized wave vectors of the transverse motion $p_n$ of the excited electron, or, in other
words, subbands $F_{+,-\frac{1}{2}}$ with different wave vectors $p_n$. Now recombination of $G$-electron and
$F$-hole is allowed by the selection rules. To conserve the total angular momentum projection an electron is
excited to $F$-subbands with $m_j=-\frac{1}{2}$, since a transition to $G$-subbands brings the small factor
$\rho$. Then, summarizing over all channels of Auger decay of the first excited biexciton states, we derive for
the AR rate
\begin{subequations}
\begin{equation}
W(T,R) = \frac{16\pi}{\hbar}e^{-\Delta E/T}\sum_n D(\delta E_n)\Gamma_n^2,
\end{equation}
where the function $D$ describes overlap of the phonon-broadened final single-exciton state of the
energy $E_x$ and initial biexciton states of the energy $E_{xx}$ with the energy detuning $\delta E = E_{xx}-E_x$.

Since at present we do not have any dependable experimental data on shapes, characteristic widths and
Stokes shifts of phonon-broadened electronic states in NWs, we use for the function $D$ a simplified
integral form $D(\delta E) = \int\text{g}(E)\text{g}(E-\delta E)dE$, assuming that the function $\text{g}(E)$
is the Lorentzian with the width $\gamma$ and the Stokes shift $\Lambda$. Then, we finally find
\begin{equation}
W(T,R)=\frac{16}{\hbar}e^{-\Delta E/T}\sum_n \frac{\gamma}{(\delta E_n-2\Lambda)^2+\gamma^2}\Gamma_n^2.
\end{equation}
\end{subequations}
To characterize the AR efficiency for multiexciton states, we introduce a ``linear'' length-independent AR coefficient
$C_L=WL^2$, which is analogous to the AR coefficient in bulk materials. Then, the AR time of multiexciton states
is found as $\tau=(C_Ln_L^2)^{-1}$, where $n_L$ is the linear density of excitons.

The results of numerical calculations of the linear coefficient $C_L$ with $\gamma = 20$ meV and $\Lambda=50$
meV for PbSe NWs of the radius $2-8$ nm at $T=300$ K are presented in Fig. 3. The total coefficient is the sum
of coefficients $C_{L_1}$ and $C_{L_2}$, which correspond to AR processes with an electron excitation to subbands
with the wave vectors $p_1$ and $p_2$, respectively.
\begin{figure}[t]
\includegraphics[width=0.87\linewidth]{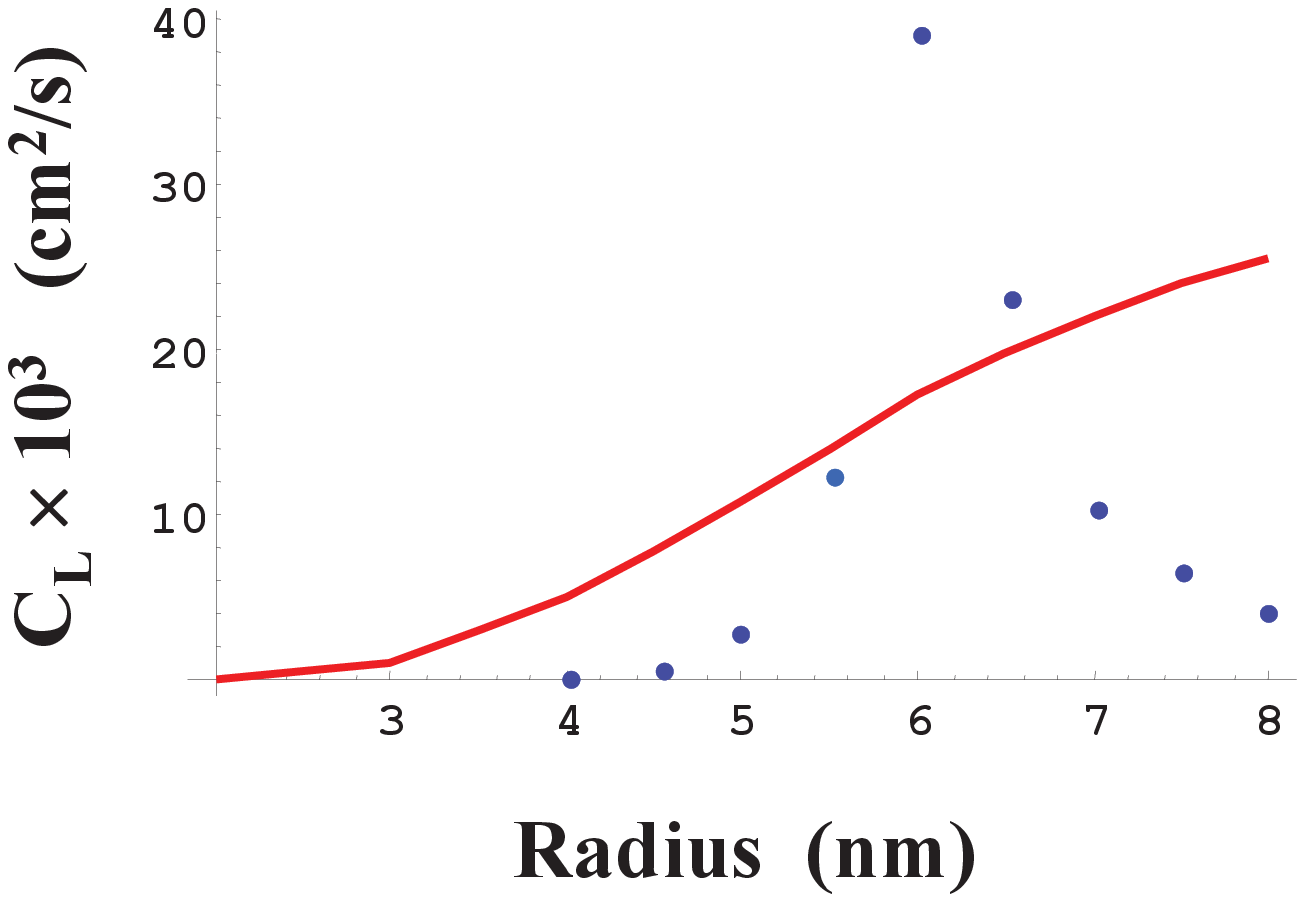}
\caption{Size dependence of the linear AR coefficients $C_{L_1}$ (curve) and $C_{L_2}$ (solid circles) in PbSe NWs of the radius
of $2-8$ nm.}
\end{figure}

Although numerical calculations are quite sensitive to unknown parameters of electron-phonon coupling, the
obtained results demonstrate basic trends in temperature and size behavior of the AR efficiency in NWs.

The matrix elements $\Gamma_1$ and $\Gamma_2$ depend on the NW radius very slowly, and size dependence of
the AR efficiency is mainly governed by size-dependent overlap of phonon-broadened single- and biexciton
states. Therefore, size dependence of the AR efficiency in NWs is reverse to that in NCs: It is greatly
reduced with decreasing $R$, while in NCs, size dependence is mainly governed by size-dependent matrix
elements of Coulomb coupling ($\propto R^{-1}$), and the AR rate scales as $R^{-3}$. \cite{K1,K2}

The coefficient $C_{L_2}$ exhibits a strong resonance behavior due to vanishing $\delta E_2-2\Lambda$ at the
radius of about 6 nm, where the phonon-broadened single- and biexciton states are well overlap each other
\cite{SEN}. Despite $\Gamma_2$ is much smaller $\Gamma_1$, $\Gamma_2\simeq 0.1\Gamma_1$, a contribution of
the resonance channel to the total coefficient $C_L$ exceeds nearby resonance a contribution of the
non-resonance channel, in which $\delta E_1-2\Lambda\gg\gamma$ at all NW radii.

To compare AR efficiencies in NWs and bulk parent materials, it is illustrative to introduce a ``volume''
coefficient $C_V=C_LS^2$, where $S$ is the NW cross-section area. The AR coefficient $C_\text{bulk}$ in
bulk PbSe was measured \cite{PbSe} to be approximately constant between 300 and 70 K with a value of about
$8\times 10^{-28}$ cm$^6$/s, and then drops a value of about $1\times 10^{-28}$ cm$^6$/s at $T = 30$ K.
The huge growth of the volume coefficient $C_V$ in NWs from $2.2\times 10^{-30}$ ($R=2$ nm; strongly
suppressed AR) to $1.2\times 10^{-25}$ cm$^6$/s ($R=8$ nm; greatly enhanced AR) is determined by increasing
both the linear coefficient $C_L$ in a wide interval from $0.14\times 10^{-3}$ ($R=2$ nm) to $30\times 10^{-3}$
cm$^2$/s ($R=8$ nm) and the NW volume.

It is easy to see that AR in NWs demonstrates a qualitatively different temperature behavior compared with
that in bulk parent material: AR in NWs is temperature suppressed due to decreasing the population of excited
biexciton states, which scales as $e^{-\Delta E_n/T}$, where $\Delta E_1$ for the first excited biexciton states
ranges in the interval from 117 meV ($R= 2$ nm) to 13 meV ($R= 8$ nm).

Overall, we have shown that owing to specific selection rules for Auger-like processes in lead-salt NWs with
a strong coupling between the conduction and the valence bands, AR is strongly suppressed in ``slim'' NWs of
the radius $R<4$ nm ($C_V\ll C_\text{bulk}$), but significantly enhanced ($C_V\gg C_\text{bulk}$) at $R>4$ nm
in comparison with bulk PbSe material. Moreover, temperature dependence qualitatively differs from that in bulk
parent material and NCs. Finally, size dependence is reversed to that in NCs: the AR rate is greatly reduced with
decreasing the NW radius, while in NCs it grows as $R^{-3}$.

Ultrafast multiexciton decay via AR is a major impediment for prospective applications of semiconductor NCs
in lasing. \cite{K2} Therefore, suppression of AR of the ground biexciton state exhibiting strong radiative
decay of electron-hole pairs with the dipole momentum mainly determined by the longitudinal Kane momentum
of bulk material makes obviously quantum-confined lead-salt NWs a subject of special interest for numerous
lasing applications in near- and mid-infrared spectral ranges.

I would like to thank Victor Klimov for numerous helpful discussions and remarks.


\begin{thebibliography}{10}
\bibitem{N}

For simplicity, we use the term ``exciton'' for an electron-hole pair in a NW, and do not account
in this Letter for bound electron-hole states.
\end{thebibliography}
\end{document}